\documentclass[10pt]{article}
\usepackage{amsmath,amsthm}
\usepackage{charter}
\usepackage{microtype}
\usepackage{graphicx}

\newtheorem{theorem}{Theorem}

\title{A radial version of the Central Limit Theorem}

\author{Kunal N. Chaudhury\thanks{kchaudhu@math.princeton.edu}}

\begin{document}
\date{}

\maketitle

\begin{abstract}
In this note, we give a probabilistic interpretation of the Central Limit Theorem used for approximating Gaussian filters in \cite{C_2010}. 
\end{abstract}

It was shown in \cite{C_2010} how a certain ``radial'' form of the Central Limit Theorem (CLT) could be used to approximate isotropic Gaussians on the plane.
The main idea was to approximate the Gaussian using box distributions that had been uniformly distributed over the half-circle. The term ``radial'' was used here to highlight 
the fact that the Gaussian filtering was achieved by convolving the image using box distributions along radial directions (and parallel to it). 

The idea was extended in two different directions in \cite{C_2010}. In one direction, it was argued that, by adjusting the widths of the box function along each 
radial direction, one could approximate anisotropic Gaussians. In particular, for the special case of the so-called \textit{four-directional box splines}, a simple algorithm was developed that allowed one
to control the covariance by simply adjusting the widths of the box distributions. In a different direction, an algorithm for space-variant filtering was developed which, unlike convolution filtering, allowed one 
to change the shape and size of the box spline at each point in the image. This was also done at the expense of just $O(1)$ operations (independent of the shape and size of the Gaussian) using a single global pre-integration
followed by local finite-differences; cf. Algorithm 1 in \cite{C_2010}.

A probabilistic interpretation of this result is as follows. Let $\boldsymbol{X}$ be random vector on the plane that is distributed on a line passing through the origin (e.g., one of the coordinate axes). Thus, $\boldsymbol{X}$ is completely specified by a probability measure $\mu(t)$ on the real line. Suppose that
\begin{equation*}
\int t \ d\! \mu(t)=0, \quad \mathrm{and} \quad \int t^2 \ d\! \mu(t)=1.
\end{equation*}
For $0 \leq \theta < \pi$, let us denote the rotation matrix on the plane by
\begin{equation*}
\mathcal{R}_{\theta}=\left(\begin{array}{cc}\cos \theta & -\sin \theta \\ \sin \theta & \cos \theta\end{array}\right).
\end{equation*}
Guided by the idea behind Theorem 2.2 in \cite{C_2010}, we can then conclude the following.

\begin{figure} 
\centering 
\includegraphics[width=1.0\linewidth]{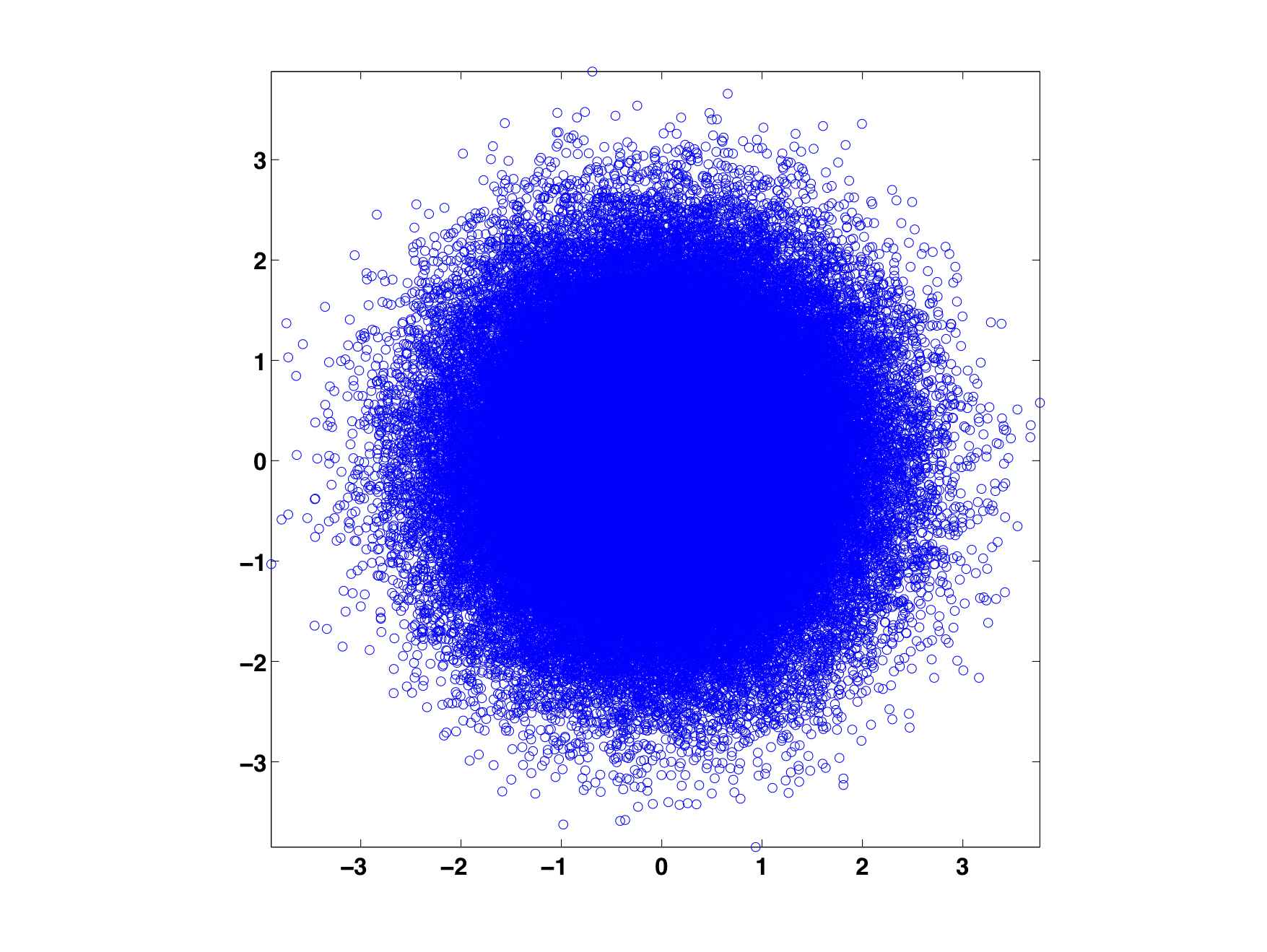}
\caption{The empirical distribution of $\boldsymbol{Z}_N$ over $10^4$ realizations when $N=10$.} 
\label{fig1}
\end{figure} 

\begin{theorem}[Radial CLT] For any integer $N$, fix $\theta_k=(k-1)\pi/N$ for $1 \leq k \leq N$. Let $\boldsymbol{X}_1,\boldsymbol{X}_2,\ldots,\boldsymbol{X}_N$ be independent and identically distributed copies of $\boldsymbol{X}$, and set
\begin{equation*}
\boldsymbol{Z}_N=\frac{1}{\sqrt N} \ \left(\mathcal{R}_{\theta_1} \boldsymbol{X}_1+\mathcal{R}_{\theta_2} \boldsymbol{X}_2+\cdots+\mathcal{R}_{\theta_N} \boldsymbol{X}_N \right).
\end{equation*}
Then the random vector $\boldsymbol{Z}_N$ converges in distribution to the standard normal distribution as $N$ gets large. More precisely, for any Borel set $B$ on the plane, 
\begin{equation*}
\lim_{N \longrightarrow \infty} \ \mathrm{Prob} (\boldsymbol{Z}_N \in B)= \frac{1}{2\pi} \int_B \exp \left(-\frac{\lVert \boldsymbol{x} \rVert^2}{2}\right) \ d\boldsymbol{x}.
\end{equation*}
\end{theorem}

Indeed, the proof of Theorem 2.2 in \cite{C_2010} shows us that that the characteristic form of $\boldsymbol{Z}_N$ converges to that of the normal distribution as $N$ gets large. 
The weak convergence of the distribution of $\boldsymbol{Z}_N$ thus follows from Levy's theorem.

We performed a simple \texttt{MATLAB} simulation for the case where $\boldsymbol{X}$ is $\sqrt {12}$ times the uniform distribution (zero mean, unit variance). The results of the simulation when $N=10$ is provided in
Figure \ref{fig1}.

\begin{figure} 
\centering 
\includegraphics[width=1.0\linewidth]{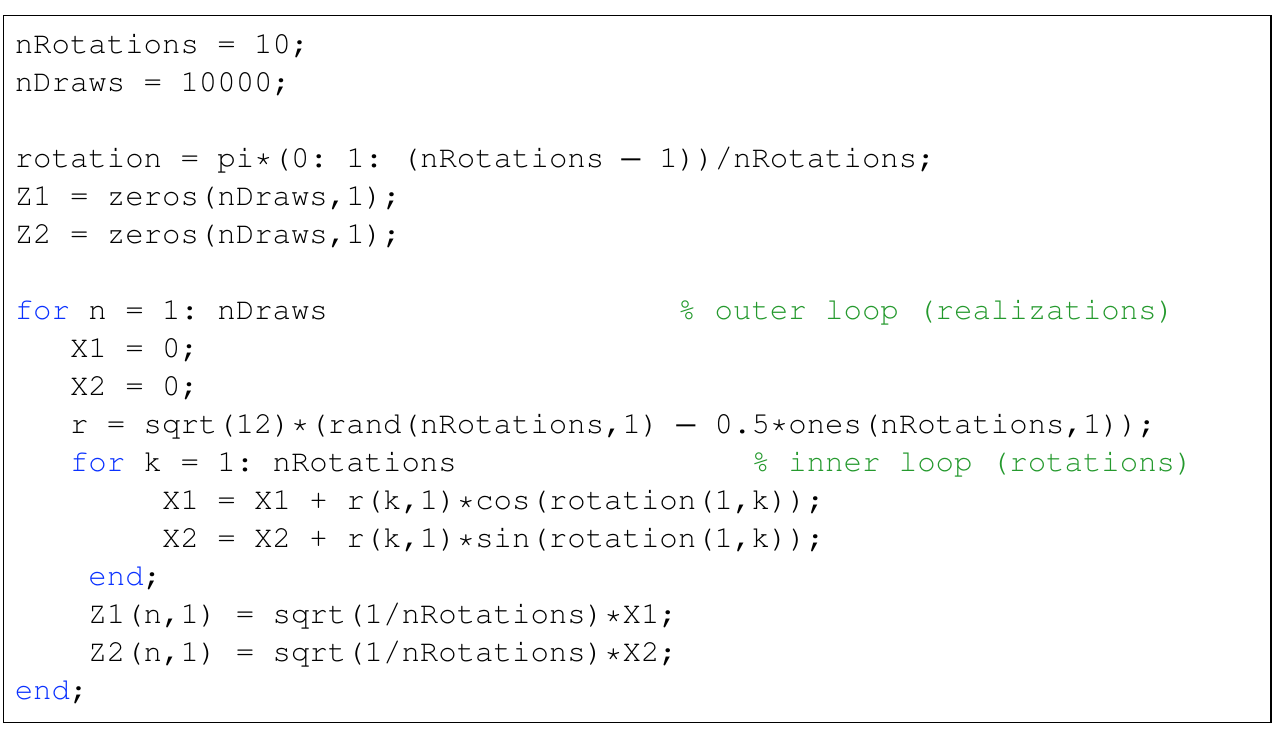}
\caption{The \texttt{MATLAB} simulation of the Central Limit Theorem.} 
\label{fig1}
\end{figure}

\end{document}